# The birth of special relativity.
## "One more essay on the subject".

Jean Reignier*

Université Libre de Bruxelles
and
Vrije Universiteit Brussel.

## Introduction.

In the general context of epistemological seminars on the occurrence of rapid transitions in sciences, I was asked to report on the transition from classical to relativistic physics at the turn of the century[1]. This paper is a revised and extended version of this report.

During the first few years of the twentieth century, two very important changes occurred in physics, which strongly influenced its development all along the century. These upheavals are: on the one hand, the irruption of the idea of quantisation, i.e. the discretisation of (some) physical quantities which up to these days were considered as continuous; and on the other hand, the relativity which deeply modified our physical conceptions of space and time. The history of the quantum revolution is well known; it has been carefully studied by many historians of science and their works do indeed converge towards a fairly unique version of the birth of the quantum theory[2]. No such general agreement seems to exist at present for the history of the birth of special relativity. The subject was treated by many authors, but their conclusions are sometimes widely divergent. Should one conclude that their approaches were not always conducted in a strictly scientific and historical way? Not necessarily. This regrettable situation could also result from different factors which can influence an "objective" analysis[3]:

- At the difference with the quantum revolution which starts at a very precise date with the work of only one physicist (i.e. Max Planck, 14 December 1900), the history of the relativistic revolution begins sometime in the ninetieth century, at a date which is more or less a matter of convention, depending on the importance one agrees to give to one physical phenomenon or another.

- The long way towards a relativistic physics is therefore much complex. Furthermore, the original texts are not always correctly appraised because we read them with the background of our relativistic scholar education.

- In modern educational systems, special relativity is generally introduced fairly early, to students which are not well informed about the problems concerning the electrodynamics of moving bodies. These problems were discussed at length during the ninetieth century; they are at the *origin* of special relativity. When modern students finally hear about them, one generally presents the sole elegant solution that relativity proposes.

- It is therefore tempting to present an oversimplified version of the history which can be caricatured as follows: "In 1905, a young genius of only 26 conceived

---

alone a wonderful theory which saved physics from the muddy situation where older conservative minds had let it go".

This presentation of the creation of special relativity by Albert Einstein (1879-1953) is readily accepted by a large majority of students, because it is simple, and pleasant to hear when one is around twenty [4]. Only a small minority notices that it contains some anomalies. Is it not strange that the fundamental relativistic transformations are called "Lorentz's transformations" and not "Einstein's transformations" (H.A. Lorentz, 1853-1928)? The perplexity of these students increases if they are told that the name was given by a third physicist (i.e. Henri Poincaré, 1854-1912), in a paper [Po-05] published one month before the famous paper by Einstein [Ei-05; 4]. Furthermore, students soon learn that the mathematical space where these transformations become fully geometric is called "Minkowski's space", from the mathematician H. Minkowski (1864-1909). One solves generally this problem of a possible profusion of creators by explaining that, if it is indeed true that Lorentz and Poincaré found some results, they were not really "relativistically minded", i.e. that they remained glued in the old vision of an absolute space filled by some mythic ether, and an absolute Newtonian time. Only Einstein proposed at once this modern vision of a relative space-time, and the mathematician Minkowski later on described its mathematical structure [Mi-09]. Of course, one should not suspect historians of science to really believe such a simplified history. Nevertheless, it has some influence, and it partly explains the definite opinion widespread among scientists about the respective merits of the creators of the new theory.

Let us give some examples of the divergent opinions that exist in the scientific literature. Let us start with the case of some well-known books:
- In 1922 appears the famous book "*Des Relativitätsprinzip* ", which presents a recollection of original papers on the special and general theory of relativity [So-22]. It was supposed to give an account of the growth of the theory, under the stimulus of physical experiment. No room is given here to the work of Poincaré [5].
- In 1951 appears the book "*A History of the Theories of Aether and Electricity* ", by Ed. Whittaker [Wi-51]. One finds here an important Chapter entitled " The relativity theory of Poincaré and Lorentz", where these two authors are presented as the incontestable creators of the relativity. Einstein's own contributions are restricted to the discovery of the formulas of aberration and of the Doppler effect.
- In 1971, M.A. Tonnelat publishes her "*Histoire du principe de relativité* " [To-71]. She explains at large that if Lorentz and Poincaré did indeed find some formulas, the "spirit" of their works was not at all relativistic, and therefore, the full merit of the creation of relativity goes to the sole Einstein.
- In 1981, A.I. Miller publishes "*Albert Einstein's special theory of relativity. Emergence (1905) and early interpretation(1905-1911)* " [Mi-81]. This book is very well documented and objective in its account of the facts. Miller presents and discusses the contributions of many authors, among them, of course, the trilogy Lorentz, Poincaré, Einstein. His conclusion is also that Einstein alone should be gratified of the "relativistic revolution". One might reproach to this otherwise excellent book, that it doesn't make any prospective effort in looking for a possible issue of the "dead" theory of Lorentz and Poincaré. Notice also that in a more recent paper [Mi-94], Miller adopts a more radical opinion: Lorentz and Poincaré could not possibly find relativity in 1905. Whatever it may be of this opinion, it remains true that the book itself is probably the best documented among the many books published up to now on the subject.
- Let me also quote the book of A. Pais "*Subtle is the Lord,...The science and life of Albert Einstein* " [Pa-82]. Chapter 3 of this book contains an analysis of the birth of special relativity which leaves the reader no doubts that the sole father of relativity is Einstein. Lorentz and Poincaré are at most precursors which didn't really *understand* the revolutionary character of the new relativistic spirit. In particular, Pais severely criticises Poincaré's habit to present Lorentz's contraction as a "third hypothesis". Pais stigmatises this attitude and considers it as an evidence that Poincaré didn't *understand* Einstein's theory. However, Pais himself does not try to analyse why, and in which circumstances, Poincaré presented relativity in that way. Alternative explanations other than a crude misunderstanding exist (see f.i. Pi-99, or Re-00).

---

[4] I have even seen a strip cartoon relating this story for children!

[5] His name appears in one of the two papers by Lorentz, but only for the remark that one should not change a theory at each new order of approximation; this remark is certainly fundamental but its creative importance is rather weak. Poincaré's paper [Po-06] is quoted twice in Notes by Sommerfeld following Minkowski's contribution, but again for unimportant details.

- I finish this rapid and limited review with the recent book of Y. Pierseaux [Pi-99]. This author proceeds to a detailed comparative analysis of Einstein's and Poincaré's approaches to relativity. He takes into account the general context in which these works were performed, and also the different attitudes of both authors with respect to physical theories. Pierseaux concludes that two slightly different theories exist, which are otherwise totally equivalent. He calls this: "*The fine structure of relativity* ".

So much for books. One finds also in the literature a lot of original papers treating the subject of the birth of relativity. Most of them do not escape the discussion of the famous question of priority. It is not possible to report here on all these papers, and I apologise for the many authors that I shall not quote. For those physicists which would like to read the fundamental paper of Poincaré with modern scientific notations and in English, I recommend the translations of H.M. Schwartz [Sc-71] and of A.A. Logunov [Lg-95]. Notice that Logunov inserts comments on Poincaré's results, and concludes to his priority. This point of view is also vigorously defended by J. Leveugle [Le-94]. Alternatively, many other authors share the opinion that Lorentz and Poincaré were working on a kind of parallel project, the relativistic character of which is at least uncertain. Therefore, they conclude that Einstein is the only true creator of special relativity. I select here (and I recognise that my choice is largely arbitrary), in chronological order and without trying to make a distinction between moderate and more radical opinions, the papers by G.H. Keswani [Ke-65], S. Goldberg [Go-67], I.Yu. Kobzarev [Ko-75], A.I. Miller [Mi-94], M. Paty [Pa-94], J. Stachel [St-95].

In view of all these divergent opinions, and keeping in mind that it will be difficult to avoid *in fine* the delicate question of priorities, I shall postpone it as much as possible and organise this paper with deep roots in the ninetieth century [6] (Parts 1 and 2). Then, I shall consider the period 1900-1910, with a particular attention to the works of Poincaré and Einstein of 1905 (Part 3). Part 4 contains my conclusions; they are rather close to the ones formulated by Pierseaux [Pi-99] who asserts that one should distinguish a kind of "fine structure" in the formulation of special relativity:

- on the one hand, a formulation with an ether which was created by Lorentz and Poincaré essentially during the years 1900-1905 from a pure electrodynamics point of view;

- on the other hand, a formulation without ether, based on Principles expressed by Einstein in 1905. This formulation was further developed by Einstein himself, but also by Planck, who became fascinated by this new physics based on Principles which renewed mechanics and thermodynamics (1906-1908). Later on, this formulation became also geometric in the hands of Minkowski (1908).

Altogether, I conclude that it would be fair to recognise the merits of at least three authors and to speak of special relativity as "the Lorentz-Poincaré-Einstein theory of relativity".

## Part 1.- Nineteenth century roots.

One can hardly deny that the deepest root of relativity is to be found in the work of Galileo who asserts that "motion is like nothing", by which he means that the description of mechanical phenomena is not affected by motion [Ga-32]. Of course, and this was promptly recognised, this assertion is only true if one restricts motion to the very special case of uniform translations. For all other kinds of motion, Newton's fundamental law implies the existence of an agent called "force". And the complete set of Newton's postulates (i.e. the fundamental dynamical law, the action-reaction principle, and S.Stevin's addition law of forces) cannot be true in arbitrary co-ordinates frames and with an arbitrary notion of time. This fact played an important role in the conviction of scientists that something like an absolute space and an absolute time should exist. However, this absolute space was considered as totally immaterial and no mechanical way exists to identify it: this is the essence of the Galilean relativity. (For a discussion of the mechanical aspect of absolute space, see for example Po-02).

---

[6] Remember that the paper was originally a contribution to seminars on the occurrence of "rapid" transitions in science. Not all these changes were really so "rapid"!

Some evidence for a "materialisation" of absolute space appears at the beginning of the ninetieth century when it was recognised that Huyghens's wave conception of light is to be preferred to Newton's corpuscular one (Thomas Young's interferences, 1801). At this epoch, a wave theory can only be thought of as a mechanical wave theory, which of course presupposes the existence of some vibrating medium. Augustin Fresnel (1788-1827) has been the champion of this idea. In a really impressive series of works, from 1815 to his death, he accumulated so much evidence in favour of the wave theory of light that the latter was later on considered as firmly established. After Fresnel, one can assert that light is just a high frequency vibration of some universal medium called ether, and therefore, physical optics can be considered as the science that studies this medium. This study revealed rather strange properties and introduced some puzzling questions concerning the nature of ether, for instance:

- the transversality of light vibration (polarisation) seems to imply a very high rigidity of ether; is this high rigidity compatible with the total lack of resistance that ether offers to the motion of matter (f.i. the motion of planets) ?

- is the ether inside transparent bodies different or alike the ether of vacuum ?

- why is the ether of transparent bodies usually dispersive, and why does this dispersive property appear to be so specific of the body ?

- what does happen to ether when a transparent body moves?

This last question will retain our attention because it played an important role in the long way to relativity. As early as 1822, Fresnel proposed a formula for a "partial" drift of the ether contained in a moving body [7]. This formula is based on his mechanical model of light propagation in ether and, following an earlier proposal by Young, on an assumed concentration of ether in transparent bodies proportional to the square of the refraction index. According to Fresnel, we get the following result. Let c be the velocity of light in vacuum and let n be the refractive index of the transparent body; according to the wave theory, the velocity of light inside the body at rest in the absolute frame of reference defined by the surrounding ether is $c/n$ . Let us now consider the same body in motion with a velocity v with respect to this absolute frame, and let us ask for the absolute velocity of light inside the moving body (for simplicity, we only consider light propagation parallel to the velocity of the body). The answer depends on the fraction of internal ether carried on by the moving body. Fresnel assumed that only the excess of ether with respect to vacuum should be considered as moving. It means that the effective velocity of body's ether is not v, but that it is reduced by the fraction $\alpha = (\rho_1 - \rho_0)/\rho_1$ , where $\rho_1$ and $\rho_0$ are the densities of ether in the body and in the vacuum, respectively. Then, according to the Galilean law of composition of velocities, the light velocity in the body as seen by an observer at rest in the absolute frame of the ether is equal to : $c/n \pm \alpha$ v , (+ or -, depending on the relative sign of the propagations, i.e. opposite or alike, respectively). If one further adopts Young's proposal of an ether concentration proportional to the square of the refractive index : $\rho_1 = \rho_0 n^2$, the "Fresnel's ether partial drift coefficient $\alpha$ " becomes :

(1)                         $\alpha = 1 - 1/n^2$ .

---

[7] The same formula was also derived later by Stokes (1846) , in a slightly different way (see [Wh-51] , Ch 4).

In the second half of the ninetieth century, several experiments were performed in order to test this value of Fresnel's drift coefficient [8] ( Fizeau 1859, Hoek 1868, Airy 1871, Michelson and Morley 1886). Fizeau's experiment is a pure local laboratory experiment which uses a high velocity water current, and directly measures the drift coefficient. All the other experiments are indirect experiments which try to exhibit the motion of the Earth with respect to ether, presumably with a velocity of the same order as its velocity around the Sun ($v \cong 30$ km/sec). They are "first order" experiments in the sense that they are sensitive to the first power of the ratio $v/c = 10^{-4}$ *if the drift coefficient has a value different of the one predicted by Fresnel's theory* . No effect of earth motion was observed, so that these experiments were interpreted as confirming the value of Fresnel's drift coefficient. I shall not report here on these experiments (see f.i. : [Wh-51] Ch 4, [To-71] Ch 4, [Re-99]). Enough is to say that the nice confirmation of Fresnel's value did endow this drift coefficient with the status of a scientific truth, of which theories had to take account.

After Maxwell's achievement of the synthesis of the electricity and magnetism theories (around 1865), it became rapidly clear that light was an electromagnetic wave phenomenon, and therefore, that one had to identify the optical ether and the electromagnetic ether (ether of Faraday). This unification solved some of the old problems of the ether theory (e.g. the transverse character of light polarisation), but at the same time it posed some new ones (e.g. should one identify the transverse Fresnel's vibration with the wave electric field or alternatively with the wave magnetic field ?). Furthermore, Maxwell's electromagnetic theory introduces phenomenological constants which are specific of the medium under consideration: it is a four fields theory (i.e. the electric field **E** , the electric displacement **D**, the magnetic field **H**, and the magnetic induction **B**), related two by two  by phenomenological constraints:

(2) $$\mathbf{D} = \varepsilon \, \mathbf{E}, \qquad \mathbf{B} = \mu \, \mathbf{H} \;.$$

The dielectric constant $\varepsilon$ and the magnetic permeability $\mu$  are empirical constants which characterise the body under study (i.e. the propagating medium). Empirically, it turns out that for higher frequency electromagnetic phenomena, like a light wave, $\varepsilon$ is not really   a constant: it depends on the frequency of the wave that propagates (dispersion). Therefore, it became evident, as already perceived in the older mechanical theory, that dispersion is to be associated with some dynamical mechanism that should be introduced in one way or another in Maxwell's theory. Notice also that Maxwell's theory says nothing about what happens when the body moves with respect to ether. There were immediately several proposals in order to take dispersion into account and to extend the theory to moving bodies, and in the matter, the question of the well established Fresnel's drift coefficient was one of the distinguishing factors. It is of course out of question to present and to discuss here these different approaches [9]. I shall limit myself to the sole Lorentz's theory which finally emerges above all others (Part 2a). On the one hand, this theory remains the basis of our present understanding of electromagnetic phenomena in matter, and on the other hand, it progressively guided Lorentz to the discovery of the relativistic transformations.

## Part 2.- Lorentz and Poincaré between 1875 and 1904.

**H.A. Lorentz.**                              **J.H. Poincaré.**
(1853-1928)                                (1854-1912)

---

[8] Fresnel's drift coefficient is also often called Fizeau's drift coefficient, because of its first experimental verification.

[9] A detailed presentation of all these tentative approaches can be found in [Wi-51].

Ph. D. at Leyden      (1875)
"Over der terugkaatsing en
breking van het licht" [10].

Engineer from Polytechnique
Paris (1875).
Doctor in Mathematics (1879) [11].

## 2a) Lorentz's classical electrodynamics.

The basic idea of Lorentz's theory is as old as the atomic idea itself: matter is nothing else than atoms moving in vacuum. The originality lies in that, on the one hand, the vacuum is now identified with Maxwell's ether which carries electromagnetic phenomena; and on the other hand, that atoms contains microscopic particles called electrons that are sensitive to the electromagnetic fields and that also contribute to create these fields. Maxwell's ether is characterised by the two electromagnetic constants $\varepsilon_0$ and $\mu_0$ which are <u>true</u> constants, totally independent of the waves that propagate in ether, and Lorentz judiciously puts them equal to one, according to an appropriate choice of units. In doing this, Lorentz effectively works with only two vacuum fields: the electric field in vacuum **e** and the magnetic field in vacuum **h**. In matter, these fields act on electrons which, by their presence and motion, act on the fields; therefore, from a macroscopic point of view, the vacuum fields are modified by the presence of matter. In that way, Lorentz succeeds to explain, at least qualitatively, nearly all phenomena concerning matter (at rest in the ether) and light: emission and absorption of light, reflection, refraction, dispersion of the refractive index, scattering of light (visible light and later X-rays), etc. [12]

Frequently also, Lorentz's qualitative explanations become quantitative provided they are completed by some phenomenological parameters in order to better characterise matter. As an example, Lorentz obtains in 1878, the celebrated Lorentz-Lorenz's formula which relates the molecular polarisability coefficient $\gamma$ to the refractive index n:

(3)               $\gamma = 3 \ (m/M) \ (n^2 - 1)/(n^2 + 2)$ ,

where m and M are the molecular and the specific masses, respectively. When used with a bit of phenomenology, this formula is very interesting: on the one hand, the atomic dispersive refraction index n can be related through Lorentz's theory to the atomic spectral lines; on the other hand, one can use the empirical additivity of the molecular polarisability of mixtures and/or of weakly bound molecules; therefore, it becomes possible to compute the refraction index of mixtures or of weakly bound molecules from the spectral lines of their atomic constituents (Cf.[Ro-65]).

At first sight, it would seem that Lorentz's model of matter with atoms bathing in an ether permanently at rest will conflict with the idea of a partial drift of the ether. But Lorentz will rapidly prove (1886) that it is not so and that his model is compatible with the *appearance*  of an ether drift coefficient, which furthermore is precisely Fresnel's one. This remarkable achievement is obtained by a reinterpretation of Maxwell's constraint relations (2) and the replacement of the static electric force e**E** which acts on electrons at rest by the "Lorentz force" which acts on electrons moving with the velocity **v** :

(3)             e **E**          $\rightarrow$           e [ **E** + (v/c) × **H** ] .

Lorentz even improves his effective drift coefficient by introducing a dispersive term which was later on observed by Zeeman (1911). More details on these questions can be found in [Wi-51], or [Re-99].

---

[10] " On the reflection and refraction of light".

[11] "Sur l'intégration des équations aux dérivées partielles à un nombre quelconque d'inconnues".

[12] A remarkable exception is the photo-electric effect which indeed needs another ingredient, i.e. Planck's quantum of action.

This Lorentz's approach of the problem of Fresnel's ether drift coefficient has a very important consequence: it proves that it is impossible to exhibit any effect of the relative motion of the Earth with respect to ether by "first order experiments". Therefore, one has to consider experiments sensitive to $(v/c)^2$, i.e. "second order experiments". At that time (1886), such an experiment existed already and it had given a negative result (A. Michelson,1881). Michelson's conclusion was that the ether was totally drifted, at the difference with Fresnel's partial drift and therefore also with Lorentz theory. However, Lorentz pointed out a small error in Michelson's calculations, which reduced the predicted effect by a factor two, so that the precision of this early Michelson experiment became insufficient to settle the question. The situation changed a few years later, when a higher precision Michelson's experiment totally confirmed the earlier result (A. Michelson et E. Morley,1887). Truly, Lorentz had to change something to his theory. He makes then the very astonishing proposal that the ether wind has a dynamical effect of contraction of all objects along the direction of motion [13] !

This contraction reduces the length of the longitudinal arm of Michelson's interferometer in such a way that it exactly compensates the second order effect of Earth motion:

(4)  $$\delta L \approx 0.5 \ L \ (v/c)^2 .$$

This reduction of length is common to all materials and is therefore not measurable by any mechanical device (length standards are reduced in just the same way). In fact, the contraction is indirectly measured by the negative result of Michelson-Morley's experiment ([Lo-95]). This is a typical case of introducing an ad-hoc explanation in order to save a theory! In the same paper [Lo-95], Lorentz introduces two other concepts which will turn out to be most important for the problem of the electrodynamics of moving bodies:

i) idea of "local time" : when considering some event which happens at time t and place (x , y, z) in a frame at rest in ether from a frame moving parallel to x with velocity v, one should not only change the space co-ordinates according to the Galilean formula:

(5)  $$x' = x - vt , \quad y' = y , \quad z' = z ,$$

but one should also use another time, the "local time", given by:

(6)  $$t' = t - vx/c^2 .$$

Indeed, it is immediately seen by direct calculation that this change of time is necessary in order to preserve the d'Alembertian [14], i.e. the basic equation of electromagnetism, up to second order terms in the ratio v/c :

(7)  $$\frac{1}{c^2} \frac{\partial^2 F}{\partial t^2} - \Delta F = \frac{1}{c^2} \frac{\partial^2 F}{\partial t'^2} - \Delta' F + O[ (v/c)^2 ] ;$$

ii) the idea of "corresponding states" : in order to obtain the invariance of the electromagnetic phenomena in the moving body frame, again up to second order terms in v/c , one has to change the electromagnetic "state" of the system (i.e. the electric and magnetic fields) in an *ad-hoc* way:

(8)  $$E'_{x'} = E_x , \quad E'_{y'} = E_y - (v/c) \ H_z , \quad E'_{z'} = E_z + (v/c) \ H_y ,$$

---

[13] The same proposal was made independently and simultaneously by G.F. FitzGerald (1892).

[14] Remember that the d'Alembertian is equal to the difference: second derivative with respect to time divided by $c^2$ minus the laplacian i.e., the sum of the second derivatives with respect to space; it is the typical equation of phenomena of wave motion (strings, sound, light, etc.)

(9)     $H'_{x'} = H_x$ ,     $H'_{y'} = H_y + (v/c) H_z$,   $H'_{z'} = H_z - (v/c) E_y$ .

Notice that change (8) is the one proposed earlier by Lorentz for the force acting on a moving electron (see Eq.3); change (9) is new.

The next year brought a triumphant confirmation of the basic ideas of Lorentz by his theoretical calculation of the recently discovered Zeeman's effect, and by the experimental verification of some subtle new details predicted by his approach. The calculation rests on the idea of a small perturbation of the classical motion of an atomic electron when one creates an external magnetic field (constant in space and time). It contains only one parameter [15], nl. the ratio e/m of the electric charge to the mass of the electron. The detailed experimental verification of Lorentz's predictions by Zeeman includes the first experimental determination of this ratio (in magnitude and sign), one year before the famous direct measurement on cathode rays by J.J. Thomson (Cf. [Lo-02], [Ro-65]). Lorentz and Zeeman will be recompensed by the Nobel Prize in Physics 1902.

We see that around the year 1900, Lorentz had succeeded to create a very powerful and flexible theory which explained all known electrodynamics effects for matter at rest in ether, and also the non observation of any new effect caused by its motion with respect to ether, including second order effects in the ratio v/c.

In reaction to critics by H. Poincaré who pointed out that the technics of changing slightly the explanation at each new order of perturbation was not really very convincing (local time in first order, contraction of lengths in second order [16]), Lorentz makes a last effort in order to produce formulae which eliminate all observable effects of motion with respect to ether to all orders in the ratio v/c. This is the famous 1904 paper: "*Electromagnetic phenomena in a system moving with any velocity less than that of light* " [Lo-04] (presented on 27 May 1904), which can be considered as the top of his work on the electrodynamics of moving bodies. One finds in this work:

   - the "correct" Lorentz's transformation [17],
   - the theorem of corresponding states written to all orders in v/c,
   - Lorentz's  formulation of the "electron dynamics". The origin of the electron mass is purely electromagnetic; as a consequence, the mass term in Newton's equations of motion depends on the velocity of the electron with respect to the ether, and it appears not to be exactly the same whether the acceleration caused by forces is longitudinal or transversal, i.e. parallel or perpendicular to the velocity; furthermore, the electron in motion with respect to ether is dynamically contracted in the direction of motion (Lorentz-FitzGerald's contraction).

In conclusion of this rapid review of the work of Lorentz, one can certainly assert the following:

   - on the one hand, Lorentz's electrodynamics is really the basis of our modern conception of matter: a microscopic approach to all phenomena, with dynamical mechanisms describing the light-matter interactions; furthermore, Lorentz was the first to recognise the necessity to bring some changes in the kinematical formulae of the Galilean transformation;

---

[15] It is a remarkable piece of chance that the quantum of action h doesn't appear in this "normal" Zeeman effect. This is not any more the case for the so-called "abnormal" Zeeman effects, where the electron spin and magnetic moment make the pattern much more complex. The abnormal Zeeman effect was only understood a quarter of a century later, after the introduction of the spin and magnetic moment of the electron.

[16] In his Lecture Notes on "Electricité et Optique" (1901), Poincaré speaks of this accumulation of hypotheses as ".. des petits coups de pouce". In "La Science et l'Hypothèse" (1902), he writes "... il fallait une explication; on l'a trouvée; on en trouve toujours; les hypothèses, c'est le fond qui manque le moins".

[17] Although it is not written in the same way as to-day. Lorentz persists in his way of thinking: he firstly performs a Galilean transformation, which he later on complements by a change of the "kinematical state", i.e.  introducing a scaling of space and of the "local time".

- on the other hand, it is not so easy to recognise Lorentz as one of the founding fathers of special relativity. He maintains the idea of an ether as a privileged medium which is the "cause" of certain *real* effects (like the Lorentz real contraction of bodies in motion with respect to ether). His transformations are basically two steps transformations: Galilean transformations (which form a group), followed by ad-hoc "corresponding states" changes (which do not). This way of reasoning forbade him to discover the group property of the full transformation, which makes his theory "relativistic". This attitude reminds us of the ancient attitude of "saving appearances" in astronomy. Once the appearances are saved, Lorentz doesn't care to re-evaluate the real role of ether.

This revaluation will be done by Poincaré and by Einstein, in two different ways however:

- Poincaré conserves the idea of an ether as a truly existing medium, because such a medium seems to be necessary for our understanding of the electromagnetic phenomena. However, he insists on the group character of the Lorentz transformations, which reduces the ether to a secondary role. The ether frames are ordinary members of the infinite set of inertial frames, and no physical experiment allows to select them in this set. Any inertial frame can legitimately ("à bon droit") be considered as the ether frame! This amounts to an elimination "*de facto* " of the ether, because we loose any possibility to identify it [Po-05], [Po-06].

- Einstein eliminates the ether "*de jure"* when he constructs a kinematics which makes no reference at all to such a medium [Ei-05; 4]. He avoids carefully to make any reference to an ether when he discusses the properties of light.

We shall analyse these two approaches in more details in Part 3.

## 2b) Poincaré's Oeuvre, from 1875 up to 1904.

Henri Poincaré is reputed as one of the most important and creative mathematicians of the last quarter of the nineteenth century. He was then considered as a kind of living King of the mathematics. His mathematical Oeuvre is immense and contains contributions to all domains of mathematics: arithmetic, geometry, algebra, analysis, ordinary and partial differential equations, group theory and analysis situs (topology). It is important for our subject to recall that Poincaré is one of the founding fathers of the theory of continuous groups (S. Lie's groups ) and in particular, that he wrote between 1899 and 1901 two important memoirs on a general presentation of this theory [18]. Poincaré was also a great mechanician: analytical mechanics, mechanics of continuous media and celestial mechanics; his work on the three body problem received the Price of the King of Sweden in 1889 and it is considered to-day as a pioneering work for modern approaches to chaos. All these mathematical tools play an important role in his works on physics, in particular in his work on relativity. To-day, it is not so well known that Poincaré was also a great physicist. This oblivion contrasts with his reputation at the beginning of the twentieth century. It is known from the archives of the Nobel Foundation that between 1901 and 1912 he was, with 49 presentations, the person most frequently proposed to the Nobel Prize in physics (Cf. Ma-98). Poincaré can be considered as the father of Mathematical-Physics, i.e. this approach to physical theories which carefully uses the richness and strength of mathematical rigor. For the period 1887-1901, his contribution to physics counts not less than (Cf. Po-01):

- 18 memoirs (1887-1892) on the differential equations of mathematical physics;
- 9 memoirs (1890-1894) on hertzian waves;
- 36 memoirs (1889-1901) critically reviewing existing physical theories;
- several books (lecture notes) corresponding to his most varied teaching.

Particularly important for our subject are his "*Théorie Mathématique de la Lumière* " of

[18] Cambridge Philosophical Transactions **18** (1899) 220-255; Rendiconti del Circolo Matematico di Palermo **15** (1901) 48 p. Completed later by a third memoir: Rendiconti del Circolo Matematico di Palermo **25** (1908) 61 p.

1899 and his "*Electricité et Optique* " of 1901; they present and discuss the various approaches to electromagnetism and optics proposed by "the successors of Maxwell".

In 1895 Poincaré publishes a series of four papers entitled "*A propos de la théorie de M. Larmor* " where he critically discusses and compares the different existing theories of electro-optics, i.e. the different adaptations of the older mechanical theories of light (Fresnel, Helmholtz, Mac Cullagh) to a Maxwellian approach, as proposed by Larmor, Helmholtz, Lorentz , J.-J. Thomson and Hertz. Poincaré proposes three criteria which he considers as being essential in order to get an acceptable theory:

1) the theory should account for Fresnel's drift coefficient;
2) it should contain the idea of conservation of electricity and magnetism;
3) it should be compatible with the Newtonian principle action=reaction.

Poincaré observes that none of the existing theories satisfies all three requirements. For example Hertz's theory violates the first condition, Helmoltz's theory violates the second, and Lorentz's theory violates the third. Poincaré discusses several alternative interpretations of this state of affairs: either the theories are incomplete, or the three criteria are (for some obscure reason) mutually incompatible, or they would become compatible only by a radical modification of admissible hypotheses. He finally concludes that one should temporarily abandon the idea to build a theory that would conform the three requirements, and that one should temporarily again retain the theory which seems *least defective* , i.e. the theory of Lorentz:

" Il faut donc renoncer à développer une théorie parfaitement satisfaisante et s'en tenir provisoirement à la moins défectueuse de toutes qui paraît être celle de Lorentz".

Nevertheless, Poincaré's disquietude with respect to the violation of the action=reaction principle is clearly stated:

"Il me paraît bien difficile d'admettre que le principe de réaction soit violé, même en apparence, et qu'il ne soit plus vrai si l'on envisage seulement les actions subies par la matière pondérable et si on laisse de côté la réaction de cette matière sur l'éther."

Finally, Poincaré makes a very important statement concerning the future of the electrodynamics of moving bodies: facts are accumulating in favour of the idea that it would turn out to be impossible to exhibit any effect of the motion of bodies with respect to ether; it seems only possible to observe effects of the relative motion of ponderable matter with respect to ponderable matter:

" L'expérience a révélé une foule de faits qui peuvent se résumer dans la formule suivante: il est impossible de rendre manifeste le mouvement absolu de la matière, ou mieux le mouvement relatif de la matière par rapport à l'éther; tout ce qu'on peut mettre en évidence, c'est le mouvement de la matière pondérable par rapport à la matière pondérable".

He suggests that one should look for a theory which fulfils this requirement to all order in v/c and he expresses the hope that the difficulty of Lorentz's theory with the principle of reaction might be solved at the same time.

As it was correctly pointed out by Goldberg [Go-67], the major importance of this paper is that it sets the framework for all Poincaré's subsequent work and attitudes in the area of the electrodynamics of moving bodies.

From now on, Poincaré will carefully watch all the developments of Lorentz's theory. At the occasion of the 25-th anniversary of the thesis of Lorentz, Poincaré contributes to the anniversary volume by a (somewhat provocative) paper where he comes back on the difficulty of the theory with respect to the principle of reaction [Po-00]. He proposes to consider as a solution that the electromagnetic energy be viewed as a kind of "fictitious fluid" with an inertial mass equal to $E/c^2$; he shows that this would be enough to restore the conservation of the momentum in the processes of emission and absorption of radiation, at least to first order in v/c . And Poincaré insists on the necessity to make use of the "local time" in order to obtain the conservation, and for the first time, he explains the physical content to this idea of local time which up to then was considered as an artificial mathematical trick:

" Pour que la compensation se fasse, il faut rapporter les phénomènes, non pas au temps vrai t, mais à un certain *temps local* t' défini de la façon suivante.

Je suppose que des observateurs placés en différents points, règlent leurs montres à l'aide de signaux lumineux; qu'ils cherchent à corriger ces signaux du temps de la transmission, mais qu'ignorant le mouvement de translation dont ils sont animés et croyant par conséquent que les signaux se transmettent également vite dans les deux sens, ils se bornent à croiser les observations en envoyant un signal de A en B, puis un autre de B en A. Le temps local t' est le temps marqué par les montres ainsi réglées.

Si alors c = 1/ √ $K_0$ est la vitesse de la lumière, et v la translation de la terre que je suppose parallèle à l'axe des x positifs, on aura:

t' = t - v x /c$^2$ . "

Poincaré states clearly that the compensation is only to first order in v/c except if one further makes another hypothesis on which he will presently not comment (clearly the Lorentz-FitzGerald contraction). It should be stressed that Poincaré was the first to discuss the concept of simultaneity and the problem of defining a common time for distant clocks. This goes back to a philosophical paper of 1898 [Po-98], but it is first explicitly borne out by a calculation in this anniversary paper of 1900. In September 1904, at the St-Louis Conference, he presents essentially the same reasoning and adds the "third hypothesis" [19] of the Lorentz-FitzGerald contraction in order to obtain the recently proposed new Lorentz's local time. In order to report on an event which takes place at point x and time t in the ether frame, one should use in the moving frame the "local time",

(10)  $\qquad t'(t,x) = \gamma ( t \ - \ vx / c^2 ) ,$

where $\gamma$ is the inverse of Lorentz-FitzGerald's contracting factor:

(11)  $\qquad \gamma = (1- v^2/c^2)^{-1/2} .$

At the occasion of this Conference, Poincaré discusses the problems met by a physics based on Principles. Among these principles, he discusses at length the Principle of Relativity which he enunciates as follows:

" The principle of relativity according to which the laws of physical phenomena should be the same, whether for an observer fixed, or for an observer carried along in a uniform movement of translation; so that we have not and could not have any means of discerning whether or not we are carried along in such a motion."

But Poincaré is too lucid a physicist to accept "Principles" as "evidently true", as Truths given by some God. Therefore he insists on the necessity to get experimental confirmation of the Principles [20].

"These principles are results of experiments boldly generalised; but they seem to derive from their generality itself an eminent degree of certitude."

Coming back to the difficulties met by the Principle of Relativity in the recent past, he concludes:

---

[19] In the past, this "third hypothesis" of Poincaré has been the source of rather radical attacks against his presentation of relativity (see f.i. Pa-82). The point was recently clarified (see Re-00). It turns out that historically, Poincaré was right to present in 1904 the Lorentz-FitzGerald contraction as a third hypothesis, and not any more after 1905. However, his reasons to leave out the third hypothesis being essentially mathematical (group arguments), they would have appeared to the physicists audiences of the time as less clear and convincing than the then generally accepted idea of a contraction.

[20] Strangely enough, this idea that Principles are largely conventions that should in any case always be borne out by experiment has been vigorously used by some authors in order to minimise the contributions of Poincaré to the theory of relativity (see f.i. Mi-94).

" Thus, the principle of relativity has been valiantly defended in these latter times, but the very energy of the defence proves how serious was the attack."

It is clear that end 1904, Poincaré is at the eve to create a new "electron mechanics". In his book on the history of relativity, Miller produces an interesting document [Mi-81]: a letter from Poincaré to Lorentz written end 1904 or early 1905 (unfortunately not dated) where he makes some points which are very illuminating of the state of evolution of the question. Lorentz's transformations contain a general scale factor (written l(v)) that Lorentz got much difficulties to put equal to one. Poincaré points out that the set of parallel translations form a group (now called the group of "boosts"), and that a rather natural hypothesis [21] on the electron structure is then enough to reduce this scale factor to one. Incidentally, this letter contains explicitly but without any comment the new relativistic formula of addition of parallel velocities [22].

# Part 3.- 1905.

The long march towards relativity is now very close to an issue. Two of the important actors of the creation of special relativity are sitting in due place. They are known as very respectable scientists and it is expected that their works (past and future) will be scrutinised by many other physicists in the world. Up to now, I did not speak much about the third main actor of the saga of relativity: Albert Einstein (1879-1955). This is not surprising since during the period covered in Parts 1 and 2, Einstein was too young to participate. Furthermore, beginning 1905, Einstein is still nearly unknown in the world of the physicists interested in electrodynamics. His early publications (five all together between 1901 and 1904) concerned thermodynamics and statistical physics. Even in this domain, these papers didn't really awake interest. Then comes the "Annus Mirabilis 1905" which will see this young physicist growing from his modest position to such a prominent one, that for many years, the whole world of physicists will pay much attention to what he says and what he writes.

It starts with a paper on the black body radiation where applying his own thermodynamical methods to examine the well known Wien's formula, Einstein shows that light exhibits features that make it alike a gas of non interacting particles. As straightforward applications of this "heuristic" point of view, Einstein gives very simple explanations of some not yet understood phenomena, the best known being the photo-electric effect. This work will be recompensed by the Nobel Prize in Physics 1921. Three other papers concern the Brownian motion and are directly derived from his Ph.D. thesis [Ei-05; 2,3,6]. They are again in the general trend of his previous work on thermodynamical and statistical physics. Even to-day, these papers remain considered as important (see comments and analysis by biographs of Einstein, f.i. Pa-82). None of these papers concern directly or indirectly the electrodynamics of moving bodies. At most can one say that his heuristic point of view on light did convince Einstein that light was not simply a wave phenomenon, and therefore, that ether might be a useless hypothesis. It is at least what Einstein himself declared in 1952 in a letter to Von Laue: " In 1905, I already knew with certitude that Maxwell's theory doesn't correctly predict the fluctuations of radiation pressure in a thermal enclosure. I convinced myself that the only way to save the situation was to give to radiation the objective status of a "being", which of course doesn't exist in Maxwell's theory".

---

[21] It is a model dependent hypothesis on the structure of the electron; it must respect the idea that the very structure of the electron can well depend on the absolute value of the velocity but not on its sign.

[22] This absence of comments was sometimes interpreted as an indication that Poincaré did not really grasp the physical meaning of this addition law (Cf. f.i. Mi-94). However, beside this implicit derivation from the group property, Poincaré makes an explicit one in his 1905 paper [Po-06] by computing the derivative of the displacement with respect to time (to local time). Furthermore, in [Po-08] he describes explicitly what does physically happen when performing such an addition. One can easily check, through a calculation following closely what he says, that he indeed describes this new addition law (Cf. Pi-99, p.143-146).

Then came the famous paper "*Zur Elektrodynamik bewegter Körper* ", [Ei-05; 4] (received on 30 June 1905), now considered by nearly all physicists as the founding paper of the special theory of relativity. As I explained in the Introduction, for many physicists this conviction is only a product of their education. Not so many of them have really read the original paper. And those who read it are stil much more numerous than the very small number of physicists who read the two papers published at the same time by Poincaré (Po-05 received on 5 June, and Po-06 received on 23 July 1905). This is certainly a very curious case in view of the respective reputation of the two scientists in 1905. One might hastly conclude that the reason is simply that one of the papers is right and the other wrong. We shall see that it is not at all so simple [23].

Let us start with a short comparison of the contents of the papers of both authors:

| **Poincaré** <br> Sur la dynamique de l'électron. | **Einstein** <br> Zur Elektrodynamik bewegter Körper. |
|---|---|
| 0- Introduction. | 0- Introduction |
|  | I- *Kinematical part*. |
| 1- Lorentz's transformation. | 1- Definition of simultaneity. |
| 2- The principle of least action. | 2- On the relativity of lengths and times. |
| 3-The Lorentz transformation and the principle of least action. | 3- Theory of the transformation of co-ordinates and times from a stationary system to another system in uniform translation relatively to the former. |
| 4- Lorentz's group. | 4- Physical meaning of the equations obtained in respect to moving rigid bodies and moving clocks. |
| 5- Langevin's waves. | 5- The composition of velocities. |
|  | II- *Electrodynamical part.* |
| 6- Contraction of electrons. | 6- Transformation of the Maxwell-Hertz equations for empty space. On the nature of the electromotive forces occurring in a magnetic field during motion. |
| 7- Quasi stationary motion. | 7- Theory of Doppler's principle and of aberration. |
| 8- Arbitrary motion. | 8- Transformation of the energy of light rays. Theory of the pressure of radiation exerted on perfect reflectors. |
| 9- Hypotheses concerning gravitation. | 9- Transformation of the Maxwell- Hertz equations when convection currents are taken into account. |

---

[23] I discard the rather trivial explanation based on the comparative fame of the periodics where these papers were published. It is true that Annalen der Physik was better known to physicists than Rendiconti del Circolo Matematico di Palermo, but the content of Poincaré's paper was previously communicated to the Comptes Rendus de l'Académie des Sciences in a sufficiently detailed way to catch the attention of physicists interested in electrodynamics [Po-05] .

This dry presentation of titles of paragraphs calls for two remarks:

i) the length of the titles is in no way representative of the real length and importance of the content of the paragraphs; remember that the total lengths of the two papers are respectively of 47 pages for Poincaré and 30 pages for Einstein;

ii) it would probably be fair to add to Einstein's paper, as a complementary eleventh paragraph, his famous paper on the equivalence of mass and energy which proceeds essentially along the same ideas (Ei-05; 5, received on 27 September 1905).

In the Introduction of their respective papers, both authors clearly announce the guide line of their works:

- Poincaré wants to continue the 1904 work of Lorentz, to put it in a more rigorous mathematical form, to discard definitely some rival models of the electron and (above all?) to try to extend Lorentz's ideas to the theory of gravitation.

- Einstein wants to eliminate from Maxwell's theory some difficulties brought in by the idea of an absolute rest. He claims that this can be done very simply with the following two postulates: the "Principle of Relativity" , and the postulate  " ... that light is always propagated in empty space with a definite velocity c which is independent of the state of motion of the emitting body." [24]

Therefore, it is immediately clear that the two authors will develop different programs.

Let us now compare in a more detailed way the contents of both papers in subdividing them according to broad subjects: Lorentz's transformation, covariance of the electromagnetism, dynamics of a relativistic particle.

1- Lorentz's transformation.

Although Lorentz's transformation is more or less present in nearly all chapters of both works, I shall limit myself here to the paragraphs that concern specifically the transformation of the co-ordinates, i.e. §1 to 5 (Kinematical part) for Einstein, and §1 and 4 for Poincaré.

In §1, Einstein discusses the notion of simultaneity and makes the very important point that this notion is only clear and evident for "local" events, i.e. events which happen at approximately the same place, and that the discussion of simultaneity of "distant" events requires at first some synchronisation of distant clocks. He proposes the synchronisation procedure by exchange of light signals, based on the second fundamental postulate of his paper that light always propagates with the same velocity c in all directions in any inertial frame of reference [25]. Since Einstein's paper does not mention any reference, one can legitimately infer that this fundamental thought is entirely of his own [26]. But one can just as well remember that Poincaré discussed the concept of simultaneity in rather similar terms as early as 1898 [27], and that he proposed

---

[24] Formulated in that way, the second postulate is a bit confusing: in a wave theory, the velocity of the wave doesn't depend on the velocity of the emitting body. Einstein means that the velocity of light has the same absolute value in all inertial systems, and that this value is independent of the direction of propagation.

[25] Einstein doesn't really define what he means by an "inertial frame of reference". He says only that it is a frame of reference where "the equations of mechanics hold good", which is a bit incoherent with his own paper. Later, in (1913), he added the footnote "to the first approximation" which of course weakens the mistake, but doesn't make the definition more precise.

[26] Some authors have discovered that in his "Bureau des Brevets" in Berne, Einstein had to examine proposals in order to synchronise the clocks of distant railways stations by exchange of telegraphic messages (private communication of I. Daubechies).

[27] This paper contains philosophical considerations on Time and its Measurement. One finds there some premonitory sentences, like the following ones: 1) about the dating of astronomical facts: "Il (the astronomer) a commencé par *admettre*  que la lumière a une vitesse constante, et en particulier que sa vitesse est la même dans toutes les directions. (..) Ce postulat ne pourra jamais être vérifié directement par l'expérience; ....". 2) In the conclusions: "Il est difficile de séparer le problème qualitatif de la

the procedure of synchronisation by exchange of light signals and its application to obtain Lorentz's local time in his 1900 paper and at the St-Louis conference of 1904 [Po-98, Po-00, Po-04]. If one can reasonably assume that both the 1898 and 1904 papers were unknown to Einstein in 1905, it is not so clear for the 1900 paper, since Einstein does refer to it in a subsequent publication of May 1906 [Ei-06]. This troublesome interrogation will probably never be answered.

In §2, Einstein discusses the application of his two Principles to two important concepts:

- the possible difference of length of a rigid rod when it is viewed from the frame where the rod is at rest, or alternatively from a moving frame;

- the delicate question of the simultaneity of distant events, viewed from their common rest frame or alternatively from a moving frame.

In my opinion, this is a most interesting paragraph, where Einstein shows most explicitly the originality of his approach.

In §3, Einstein proceeds to a derivation of Lorentz's co-ordinates transformation from the two enunciated Principles. Notice that one doesn't find the name of Lorentz in this paragraph, nor the name of any one else who wrote or used the same formulae before 1905 (Voigt 1887, Lorentz 1899 and 1904, Larmor 1900, Poincaré 1904). This is a bit strange because these formulae were already well known. This silence can be put in parallel with the complicated title of the paragraph; it is possible that Einstein wants to emphasise the difference between *his interpretation* and the one of previous authors. The message may be: *the physical content of these formulae being new, the formulae are new*. Here again, we shall probably never know. Schematically, Einstein's demonstration proceeds along three steps:

- The first step is a detailed analysis of the events corresponding to a go and back light exchange along the direction of propagation of the moving frame, with the hypotheses that the transformation is linear and that the light velocity is the same in both directions in all frames. This gives him the ordinary Lorentz's transformation, including the arbitrary global scale factor l(v) that we mentioned earlier; Einstein writes it $\phi(v)$. It is interesting to remark that Einstein does separate this factor. This is a kind of a priori choice which has no real justification, except of course if one already knows the answer (again the same interrogation about a previous knowledge). He meets then the old problem of how to get rid of this factor $\phi(v)$.

- In the second step , Einstein performs a kind of inverse transformation, and deduces from it that :

(12)                                     $\phi(v) \, \phi(-v) = 1$.

In essence, this is a first application of the principle of relativity, with the requirement of a total reciprocity between the two translating frames.

- In the third step , Einstein considers again the events corresponding to a go and back light exchange but now in a direction perpendicular to the translation. This is entirely new! I mentioned already that Poincaré had discussed several times the synchronisation of clocks placed along the direction of translation (i.e., essentially Einstein's first step). But Einstein is the first to complement this longitudinal synchronisation by a transverse one. He deduces from this operation the physical meaning of the factor $\phi(v)$: it corresponds to a contraction of the length of a transverse rigid rod when it is seen from a moving frame. He makes then the clever statement that, simultanéité du problème quantitatif de la mesure du temps; soit qu'on se serve d'un chronomètre, soit qu'on ait à tenir compte d'une vitesse de transmission, comme celle de la lumière, car on ne saurait mesurer une pareille vitesse sans *mesurer* un temps." 3) Again in the conclusions: "Nous n'avons pas l'intuition directe de la simultanéité, pas plus que celle de l'égalité de deux durées. (....) La simultanéité de deux événements, ou l'ordre de leur succession, l'égalité de deux durées, doivent être définies de telle sorte que l'énoncé des lois naturelles soit aussi simple que possible."

because the rod is perpendicular to the direction of motion, this possible contraction can well depend on the relative velocity, but not on its sign! Therefore:

(13) $$\phi(v) = \phi(-v),$$

which, together with (12), gives:

(14) $$\phi(v) = 1.$$

Einstein's derivation of Lorentz's transformation is most interesting in that it differs on some important points from the previous approaches by Lorentz and by Poincaré. Firstly, Einstein constructs a *proof* of Lorentz's transformation from *first principles*. Secondly, he discusses the relations between *two inertial frames*, none of them being privileged; this is particularly clear in step 2. Contrariwise, Lorentz and Poincaré always discuss the relations between the privileged ether frame and another frame in uniform translation with respect to the former. At first, it seems that the latter attitude is not "relativistic". However, it has some advantage because it avoids the delicate question of defining the concept of inertial frame: there exists an ether, and the frames at rest in ether are absolute Newton's frames. Inertial frames are then frames in uniform translation with respect to the ether.

For Lorentz, in full conformity with this idea, the fundamental transformations remain the Galilean transformations, and the rest of the story is a question of changing things in an appropriate manner (local time, scaling, corresponding kinematics and electromagnetic states), in order to obtain a *formal* appearance of equivalence of the electromagnetic phenomena in moving frames and absolute frames.

For Poincaré, it is not quite so simple. His approach evolves slowly from the original Lorentz's position towards a detailed study and possibly a *proof* of Lorentz's transformation, and a revision of the status of the ether. Indications and evidences of this evolution are:

- Firstly, that the calculations giving the physical meaning of Lorentz's local time (1900, 1904) are hardly separable from some implicit demonstration of Lorentz's transformation, similar to Einstein's first step (see Re-00 for details).

- Secondly, the evolution in time of Poincaré's attitude about the problem of the scale factor (i.e. Lorentz's l(v) or Einstein's $\phi(v)$). In his letter to Lorentz, he recognises the existence of the group of "boosts", and he finds the relativistic addition of parallel velocities and a multiplication law for the scale factor: two successive parallel boosts characterised by velocities v and v' are equivalent to a parallel boost characterised by a velocity v'', with the rules:

(15) $$v'' = (v + v')/(1 + v\,v'/c^2),$$

(16) $$l(v'') = l(v)\,l(v').$$

At the time of his letter to Lorentz, Poincaré doesn't yet use fully the resources of group theory, so that he comes back to the electron theory where he picks up a currently used explicit form of l(v):

(17) $$l(v) = [1 - (v/c)^2]^m,$$

where m is some model dependent parameter; clearly, only m=0 can fit the multiplication law (16), and this means that l(v) is equal to one. But in §4 of his 1905 paper, Poincaré goes much farther. He makes a complete analysis of the Lorentz's group (so did he term this new group), i.e., including not only the boosts, but also the spatial rotations. Then, he can easily show without any hypothesis coming from the electron theory, and without the transverse synchronisation, that l(v) must be equal to one.

- Thirdly, this group structure being firmly established and remembering that Poincaré knows very well what a group structure means, it becomes evident that the ether frames loose their privileged status. This understanding is clearly stated in the introduction of the paper:

".."; deux systèmes, l'un immobile, l'autre en translation, deviennent ainsi l'image exacte l'un de l'autre."

Let us also remark that Poincaré's study of the Lorentz's group is extraordinary modern:
- derivation of the associated Lie group;
- theorem that any Lorentz's transformation can be seen as a Lorentz's boost along x, preceded and followed by an appropriate rotation;
- theorem that any Lorentz's transformation can be resolved into a dilatation and a linear transformation which leaves unaltered the quadratic form: $x^2 + y^2 + z^2 - t^2$ ;
- geometrical interpretation of the latter (continuous) transformation as a rotation in the four dimensional space:

$$x , y , z , t \sqrt{-1} ;$$

- discovery of the electromagnetic field invariants (§ 3 , 5):

(18)     $\mathbf{E}^2 - \mathbf{H}^2,$          $\mathbf{E} . \mathbf{H}$ ,

and of several kinematics invariants when more particles are present (§9);
- proof that several physical quantities are the individual components of "four partners" that vary under Lorentz's transformations like the three space co-ordinates and the time. Examples are: the force reduced to unit volume or (alternatively) reduced to unit charge and the corresponding work per unit time, the four component velocity : $\gamma \, \mathbf{v}, \gamma$  (or momentum: $m\gamma\mathbf{v}, m\gamma$ ; as usual, I write $\gamma = (1- (v/c)^2)^{-1/2}$ ), the electric current  and electric charge densities, the vector and the scalar electromagnetic potential in Lorentz's gauge, etc.

Let us finally notice that the new law of addition of velocities (parallel and non-parallel composition) was found by Einstein (§5) and by Poincaré (§1) in nearly the same way: the combination of the velocity vector $\mathbf{w}$ of a particle with the velocity v of a frame moving along the x axis. Einstein doesn't hesitate to extrapolate the law to the case of a "light particle" moving along x with the velocity c and he finds in this calculation a confirmation of his postulate that the light velocity remains equal to c in all inertial frames [28]. This kind of extrapolation will of course never be done by Poincaré who considers that light and material particles have completely different status.

2- Covariance of electromagnetism.

Let us now consider the electrodynamical part (except the dynamical equation for particle motion, which we will discuss later), i.e. § 6-9 for Einstein, and § 1-3 and 5 for Poincaré. Both authors want to make sure that Maxwell's equations are invariant under Lorentz's transformation. In vacuum, where $\varepsilon_0$ and $\mu_0$ can be taken equal to one (see Eq.2), these equations can be written in two equivalent ways:

- Either , in terms of the physical fields, electric field $\mathbf{E}$ and magnetic field $\mathbf{H}$ (remember that $\varepsilon_0$ and $\mu_0$ are equal to one):

(19)          div $\mathbf{H}$ $= 0$ ,

---

[28] Funny enough, Einstein doesn't make w = c but v = c ! This amounts to consider a Lorentz's transformation of velocity c, i.e. a frame (an observer) going as fast as light, clearly an impossible case. Simple misprint or Einstein's subconscious old dream of following a light ray ?

(20) $\qquad$ $\text{rot } \mathbf{E} + \dfrac{1}{c} \dfrac{\partial \mathbf{H}}{\partial t} = 0 \ ,$

(21) $\qquad$ $\text{div } \mathbf{E} = 4 \pi \rho \ ,$

(22) $\qquad$ $\text{rot } \mathbf{H} - \dfrac{1}{c} \dfrac{\partial \mathbf{E}}{\partial t} = 4 \pi \mathbf{j} \ ,$

where $\rho$ is the electric charge density and $\mathbf{j}$ is the electric current density; these are obviously constrained by the relation:

(23) $\qquad$ $\text{div } \mathbf{j} + \dfrac{\partial \rho}{\partial t} = 0 \ ,$

which expresses the conservation of electricity; in the case of a material current with a "local" charge of velocity $\mathbf{u}$, the density current $\mathbf{j}$ is equal to $\rho \, \mathbf{u}$;

  - <u>Or</u> , in terms of a scalar potential V and a vector potential $\mathbf{A}$ constrained by a gauge condition; this corresponds to a trivial integration of the homogeneous equations (19) and (20):

(19') $\qquad$ $\mathbf{H} = \text{rot } \mathbf{A} \ ,$

(20') $\qquad$ $\mathbf{E} = - \dfrac{1}{c} \dfrac{\partial \mathbf{A}}{\partial t} - \mathbf{grad} \, V \ ,$

(21') $\qquad$ $\dfrac{1}{c^2} \dfrac{\partial^2 V}{\partial t^2} - \Delta \, V = 4 \pi \rho \ ,$

(22') $\qquad$ $\dfrac{1}{c^2} \dfrac{\partial^2 \mathbf{A}}{\partial t^2} - \Delta \, \mathbf{A} = 4 \pi \mathbf{j} \ ,$

with Lorentz's gauge constraint:

(24) $\qquad$ $\text{div } \mathbf{A} + \dfrac{1}{c} \dfrac{\partial V}{\partial t} = 0 \ .$

  It is interesting to notice that Einstein chooses to use the first set of equations while Poincaré chooses to use the second [29]; furthermore, the way they respectively proceed is characteristic of their preoccupations. I shall not reproduce here the calculations of these authors, but I shall try to indicate their ways of reasoning.

  One remembers that Poincaré insisted in his paper "*À propos de la théorie de M. Larmor* " of 1895 , on the absolute necessity of the conservation of the electric charge. He will now apply it on the changing of frame operation: the total electric charge in some spatial volume in the (x,y,z,t) frame must be the same as the total electric charge in the corresponding volume (obtained through Lorentz's transformation) in the (x', y', z', t') frame. In that way, he gets the relation between corresponding electric charge densities $\rho$ and $\rho'$. Since the kinematics gives the relation between the velocities $\mathbf{u}$ and $\mathbf{u}'$, he immediately obtains the relation between the electric current densities $\mathbf{j}$ and $\mathbf{j}'$. He checks the covariance of the conservation law (23) [30]. He also checks the formal covariance of the d'Alembertian operator and, imposing the covariance of Eqs. 21' and 22' , he gets the relation between the four partners $A_x$ , $A_y$ , $A_z$ , V and their homologues $A'_{x'}$, $A'_{y'}$ , $A'_{z'}$ , V' ; he sees that they transform just alike the co-ordinates x , y , z , ct (four-vectors) ; he checks the covariance of the gauge constraint (24). From the kinematics, the latter results, and Eqs. 19' and 20' , he computes the relations between the six partners $E_x$ , $E_y$ , $E_z$ , $H_x$ , $H_y$ , $H_z$ , and their homologues $E'_{x'}$ , $E'_{y'}$ , $E'_{z'}$ , $H'_{x'}$ , $H'_{y'}$ , $H'_{z'}$ , as they were already found by Lorentz [Lo-04] [31]. He finally checks the covariance of the Eqs. 19 to 22. Poincaré considers then the Lorentz's force on the unit electric charge, when this force is written in just the same formal way in both frames; he shows that there exists a "four partners" set : $F_x$ , $F_y$ , $F_z$ , **F.u** , that transforms just in the same way as the co-ordinates x , y , z , ct (four vector). This achieves the proof of covariance of the Maxwell equations.

---

[29] Poincaré's results are obtained for the full conformal group, including the dilatation parameter. Only later, does he prove that l(v) is equal to one. In this comparative report, I take l(v) =1 from the beginning.

Einstein considers firstly equations (20) and (22) with **j** = 0, i.e. the electromagnetic field in empty space (no electric charge and no electric current). He takes these equations as they are written in one frame, performs Lorentz's transformation on the co-ordinates and assumes that the so obtained equations have again the same form (20) and (22) in the other frame. This gives him the relations between the six partners $E_x$ , $E_y$ , $E_z$ , $H_x$ , $H_y$ , $H_z$ , and their homologues $E'_{x'}$ , $E'_{y'}$ , $E'_{z'}$ , $H'_{x'}$ , $H'_{y'}$ , $H'_{z'}$ ; these relations are those already found by Lorentz [Lo-04]. Then, he immediately looks for some physical consequences of this transformation law of the free-fields (§ 7 and 8). In that way, he obtains the relativistic formula for the Doppler effect (including the entirely new transverse Doppler effect), the aberration of light, the radiation pressure on perfect reflectors, and the transformation of the energy of a "light complex" [32] . He notes (without any allusion to the famous quantum formula E = hν that he derived three months earlier):

"It is remarkable that the energy and the frequency of a light complex vary with the state of motion of the observer in accordance with the same law."

In § 9, Einstein terminates his proof of the covariance of Maxwell's equations. He considers Eqs. 20-22 including now the electric charge density ρ and the electric current density ρ **u**. Assuming the equations to be valid in one frame [33], he performs the co-ordinates transformation and the transformation of the fields that he found in § 6 for the free fields. This latter operation indicates that Einstein considers the fields (and therefore the light) as physical entities, which should transform in the same way, independently of their possible coupling to charges and currents. From these transformations, he deduces the transformation law of the densities of electric charge and electric current. He notes (with an evident satisfaction) that the latter is only covariant if one adopts the new law of addition of velocities that he found in § 5:

---

[30] Poincaré's results differ slightly from Lorentz's ones. Only Poincaré's results do satisfy the covariance of the conservation law (23).

[31] As it is well known, they only differ from the older Lorentz's formulae [Lo-95] reproduced in our Eqs. 8 and 9 by the presence of a factor γ in the right hand side of the components $E'_{y'}$, $E'_{z'}$ , $H'_{y'}$ and $H'_{z'}$ .

[32] Einstein's conceptual relations with light are not simple: he speaks alternatively of "light rays", "light waves", and of "light complex", the latter remaining undefined. In the case of the energy, it seems that the "complex" is the light contained in some sphere moving with the wave (with velocity c) in one frame. This sphere is viewed as an ellipsoid moving with velocity c in the other frame. The energy of the "complex" is the product of the "time averaged energy density " times the volume. Notice that if Einstein discusses at length properties of "existing light" (Doppler effect, energy of a "light complex", radiation pressure), he avoids to discuss the process of emission of light by accelerated particles.

[33] For the first time, Einstein makes here an allusion to Lorentz : "If we imagine the electric charges to be invariably coupled to small rigid bodies (ions, electrons), these equations are the electromagnetic basis of the Lorentzian electrodynamics and optics of moving bodies" . This doesn't tell us much on which work of Lorentz between 1886 and 1904 he refers to.

"Since - as follows from the theorem of addition of velocities (§ 5) - the vector **u**' is nothing else than the velocity of the electric charge, measured in the system k, we have the proof that, on the basis of our kinematical principles, the electrodynamic foundation of Lorentz's theory of the electrodynamics of moving bodies is in agreement with the principle of relativity."

The next sentence is interesting; Einstein writes:

"In addition, I may briefly remark that the following important law may easily be deduced from the developed equations: If an electrically charged body is in motion anywhere in space without altering its charge when regarded from a system of co-ordinates moving with the body, its charge also remains - when regarded from the "stationary "system K - constant."

In other words, the final statement of these paragraphs on electrodynamics is just the starting point of the analogous Poincaré's study. This underlines the difference of point of view between the two authors in their simultaneous and original first complete proofs of the covariance of Lorentz's electrodynamics.

Let me now come back on § 5 of Poincaré's paper (Langevin waves) which has no counter part in Einstein's paper. It concerns the electromagnetic field created by a charged particle in motion. It is well known that this field can be separated into two parts: a part linear in the acceleration of the particle (that Langevin calls "the acceleration wave") and another part which depends only on the velocity (that Langevin calls "the velocity wave"). We know that only the former subsists asymptotically and that it represents the radiation (light) actually emitted by the accelerated particle. Poincaré observes that the velocity wave is nothing else than the Lorentz's transform of the static electric field of the particle at rest, when the particle is given its velocity by a proper Lorentz's transformation. Things are not so easy to handle for the accelerated wave. However, Poincaré remarks that essential features of the wave, previously discovered by Hertz for the emission of radiation by a slowly moving accelerated particle, must remain true when the particle has a higher velocity because they correspond to *Lorentz's invariant quantities* . These properties are the following ones: the electric and magnetic fields have the same magnitude, they are mutually perpendicular and they are both perpendicular to the normal to the wave front. Poincaré shows here his handiness in manipulating invariants to elucidate rather complicated physical phenomena.

But Poincaré is not yet completely satisfied with his direct check of the covariance of Maxwell-Lorentz's equations of §1. He wants to complete his analysis by a careful study of the derivation of these equations from a principle of least action (§ 2 and 3). This part of the paper is really a master piece of the calculus of variations. Poincaré succeeds to derive Maxwell-Lorentz's equations and Lorentz's force in a covariant way from the variation calculus. He uses the following "action" [34]:

$$(25) \qquad J = \int dt\, d\tau \left[ \left( \frac{\mathbf{E}}{t} + \rho\, \mathbf{v} \right) . \mathbf{A} - \frac{1}{2} \left( \mathbf{E}^2 + \mathbf{H}^2 \right) \right] ,$$

the integral being taken over all space ($d\tau$ is the volume element of space ) and time; alternatively, he also uses the simpler form:

$$(25') \qquad J = \int dt\, d\tau\, \frac{1}{2} \left[ \mathbf{H}^2 - \mathbf{E}^2 \right] ,$$

which is equivalent to the former and which will prove easier to handle in the calculations of the dynamics of the electron.

---

[34] Poincaré's calculus is based on conventions which differ by signs from the modern notations. Therefore, I change the sign of his lagrangian density in order to agree with current conventions.

Furthermore, as it was correctly pointed out by Logunov [Lg-95], some details of the calculations of § 3 leave no doubt about the understanding by Poincaré of the relativity of length and time intervals measured in different inertial reference frames.

## 3-The dynamics of the electron.

Let us now consider the problem of establishing the dynamical equations of motion for a particle. No other part of the detailed comparison of the works of both authors can better illustrate the difference of their interests. Einstein's treatment of the problem is short (§10), simple and totally unusual for the epoch. He doesn't refer to any kind of electromagnetic structure of the electron as it was then customary (Cf. the works of J.J. Thomson (1881), J. Larmor (1894), Abraham [Ab-02], Lorentz [Lo-04], Langevin [La-05] and others). Poincaré's treatment is extensive (§ 6-8), of higher mathematical level, and essentially based on the electromagnetic model of the electron; it also provides the title of the paper.

Let us start with Einstein's treatment of the problem. He considers some material particle with a definite mass and electric charge, and he calls it "electron" only for the sake of convenience. Einstein assumes that, if this particle is instantaneously at rest, then Newton's dynamical equation of motion must be valid, with the sole electric force on the right hand side. According to the principle of relativity, if he then communicates to this particle its actual velocity by an appropriate Lorentz's transformation, transforming of course accordingly the acceleration and the force, he finds the relativistic instantaneous equation of motion. Einstein performs this transformation for an instantaneous velocity along the x-axis and stops there his reasoning. He apparently doesn't realise that this equation is then only instantaneously valid and that an instant later, the motion will not any more be along the x-axis. Therefore his dynamical equation of motion is not correct. Furthermore, probably influenced by Abraham's work (not quoted), he makes a wrong choice for the definition of the force and accordingly, he gets a wrong value for the "transverse" mass. Nevertheless, he mentions that:

"With a different definition of forces and accelerations we should naturally obtain other values for the masses. This shows us that in comparing different theories of the motion of the electron we must proceed very cautiously."

He generalises his results by a mere sleight of hand:

"We remark that these results as to the mass are also valid for ponderable material points, because a ponderable material point can be made into an electron (in our sense of the word) by the addition of an electric charge, *no matter how small* ."

It looks as if Einstein were not really interested in the question of the electron dynamics [35]. Nevertheless, he goes one step further with his (wrong) equation and he uses (correctly!) the sole x-component for a finite rectilinear motion along the x-axis, in order to compute the kinetic energy of the electron as being the energy withdrawn from the electrostatic field when the electron increases its velocity from zero to its actual value v. In that way, he finds that the kinetic energy of the electron is equal to:

(26) $$W = mc^2 \left[ (1 - v^2/c^2)^{-1/2} - 1 \right] .$$

This formula interests him much more than the dynamical equations themselves. He extends immediately its validity to any ponderable matter, ".... by virtue of the argument stated above ". We shall see later how this bold extension will allow him to derive his famous mass-energy equivalence formula.

One sees that Einstein did not really succeed in 1905 to get the correct dynamical equations of motion of an electron. But his genial intuition was then enough to put him right away very close to his most extraordinary discovery of the equivalence of mass and energy [Ei-05; 5].

---

[35] The correct dynamical equations of motion were obtained along Einstein's way of reasoning by Max Planck [Pl-06]. See also Pl-07 and Pl-08.

Let us now come to Poincaré's treatment of the problem. I shall present it with some details, because it is really the centre of the paper. Poincaré adopts fully the electromagnetic image of the electron, as it was previously proposed by several authors (see refs. here above): dynamically, the electron is nothing else than the electromagnetic field created by its own electric charge. Therefore, the "action" that defines the electron free motion is the electromagnetic action (25'), where the fields are created by the charge e. In order to avoid the divergence of the action integral at small distances that happens in the case of a point particle, one has to assume that the charge is distributed in some (unknown) way in a very small volume. All authors agree that, in the absence of any physically privileged direction (i.e., for an electron at rest in the ether), this distribution of charge is spherically symmetric. The models differ from each other when some privileged direction exists, f.i. when the electron is in uniform translation with respect to the ether. Poincaré wants to compare these different models on grounds of their reaction to the general Lorentz's transformation, including the scale factor l(v). Beside the traditional cases of the electron at rest in the ether and the electron in uniform translation with respect to the ether, he wants to consider what happens to the latter when one *formally* carries it to rest by a Lorentz's transformation; this is the essence of his distinction between the "real" electron (l'électron "vrai") and the "ideal" electron (l'électron "idéal"). He adopts a general ellipsoidal model with cylindrical symmetry around the velocity, therefore characterised by two parameters: the longitudinal axis r, and the transversal axis q r. As usual the electron is spherically symmetric when at rest in ether, but this hypothesis turns out to be of no importance in the following, because the discussion focuses on the difference between the "real" electron and the "ideal" electron. Poincaré pays special attention to the three following models, but his discussion remains general:

| Electron model. | At rest in ether. | Real electron. | Ideal electron. |
|---|---|---|---|
| Abraham: | Spherical and rigid. | | Prolate ellipsoid. |
| Langevin: | Spherical. | Oblate ellipsoid, constant volume. | Spherical, but dilated volume. |
| Lorentz: | Spherical. | Oblate ellipsoid, contracted volume. | Spherical, alike in the ether. |

If one compares these three models on grounds of the general conformal Lorentz's transformation with l = l(v), one sees that Abraham's model is not compatible with this transformation (rigid spherical electron), that Langevin's model corresponds to l(v) = [1 - (v/c)$^2$]$^{1/6}$, and that Lorentz's model corresponds to l(v) =1. If one extends the comparison within the larger class of models considered by Poincaré, the conclusion remains that Lorentz's model is *the only one* compatible with l(v) =1, i.e. compatible with a group structure based on the sole relative velocity. In other words, Lorentz's model is the only one which can be called "relativistic". In the crucial choice which he will have to make, between a very satisfactory purely electromagnetic model which unfortunately is non relativistic (Langevin), and a non completely electromagnetic model which is relativistic (Lorentz), Poincaré doesn't hesitate:

"L'avantage de la théorie de Langevin, c'est qu'elle ne fait intervenir que les forces électromagnétiques et les forces de liaison: mais elle est incompatible avec le postulat de relativité; c'est ce que Lorentz avait montré, c'est ce que je retrouve à mon tour par une autre voie en faisant appel aux principes de la théorie des groupes. Il faut donc en revenir à la théorie de Lorentz; mais si l'on veut la conserver et éviter d'intolérables contradictions, il faut supposer une force spéciale qui explique à la fois la contraction et la constance de deux des axes . J'ai cherché à déterminer cette force, j'ai trouvé qu'*elle peut être assimilée à une pression extérieure constante, agissant sur l'électron déformable et compressible, et dont le travail est proportionnel aux variations du volume de cet électron* ".

The choice is clear and shows the importance that Poincaré gives to the principle of relativity.

The calculations of Poincaré are too complex to be reproduced here. Nevertheless, it is important to give some details in order to appreciate the crucial choice that is made here. Starting with the ordinary Maxwell's expressions for the energy and the momentum of a purely electromagnetic system, Poincaré computes these quantities for the ideal electron (W', $\mathbf{P'} = 0$) and for the real electron (W, $\mathbf{P}$). With the help of Lorentz's transformation of the fields and of the volume element dt , he finds:

$$(27) \qquad W = \frac{W' \; l(v)}{\sqrt{1 - (v/c)^2}} \left( 1 + \frac{1}{3} (v/c)^2 \right) ,$$

$$(28) \qquad P_x = \frac{4}{3 \, c^2} \; \frac{W' \; l(v)}{\sqrt{1 - (v/c)^2}} \; v ,$$

(the components $P_y$ and $P_z$ being zero for a real electron in uniform translation along the x axis). Poincaré computes also the Lagrangian function (i.e. the action by unit of time) of the electromagnetic electron from Eq. 25' :

$$(29) \qquad L = \int d\tau \; \frac{1}{2} \left[ \mathbf{H}^2 - \mathbf{E}^2 \right] = - W' \; l(v) \sqrt{1 - (v/c)^2} \; .$$

When he compares the canonical expressions of energy and momentum of a particle derived from this Lagrangian with the values (27) and (28), he sees that the full scheme is consistent only if $l(v) = [1 - (v/c)^2]^{1/6}$, i.e., the value of Langevin's model. A simple check of this calculation (otherwise, a bit sophisticated) is obtained by computing the low velocity approximations of (27) and (28):

$$(27') \qquad W = W' + \frac{W'}{c^2} \left( \frac{5}{6} + l'(0) \right) v^2 + O(v^4) ,$$

$$(28') \qquad P_x = \frac{4}{3} \; \frac{W'}{c^2} \; v \; + \; O(v^3) \; ,$$

from which we see that the masses associated, on the one hand, with the kinetic energy, and on the other hand, with the momentum, are only equal if the derivative $l'(0) = - 1/6$ . This is the case of Langevin's model. But Poincaré rejects Langevin's model because it is non relativistic, and he takes the alternative option to complete Lorentz's model by introducing non electromagnetic forces. Through rather cumbersome calculations, he finds that it is sufficient to introduce a constant external pressure whose work is proportional to the volume of the electron [36]. It is remarkable that this complementary potential of forces doesn't change the general form of the Lagrangian (29), now written with l(v) =1. One gets in that way the full compatibility of the general canonical formalism with the principle of relativity. The rest of the story is then straightforward: from the Lagrangian (29') of the electron in free motion [37], the canonical formula (30), and the Lorentz's force acting as an external force on this electron, one obtains Poincaré's dynamical equation (31) :

$$(29') \qquad L = - W' \sqrt{1 - (v/c)^2} \ ,$$

$$(30) \qquad \frac{d}{dt}\left(\frac{L}{v_k}\right) = F_k \ , \ (k = 1,2,3) \ ,$$

$$(31) \qquad \frac{d}{dt}\frac{(W'/c^2)\,v_k}{\sqrt{1 - (v/c)^2}} = e\,[\,E_k + (\mathbf{v} \times \mathbf{H})_k\,] \ , \ (k = 1,2,3).$$

---

[36] Nowadays, with our better knowledge of the relativistic tensor calculus, the calculations of Poincaré can be understood in a much simpler way. The energy-momentum tensor $T\mu\nu$ defines an energy-momentum four vector $P_\mu$ by integration of the fourth components $T\mu0$ over all space, if and only if, it is conserved: $\partial_\nu T\mu\nu = 0$. This is not possible for the electromagnetic energy momentum tensor in case of the presence of electric charge (the electron). Therefore, one has to complete this electromagnetic energy-momentum tensor in order to satisfy the conservation condition. The simplest solution is to introduce ad-hoc diagonal terms, like a constant external pressure for the space components $T_{kk}$ , and the corresponding work for the time component $T_{oo}$. (Cfr. for instance: Ar-66 or Jc-75).The pressure and its work can eventually be computed if one chooses a model for the electrostatic charge distribution of the electron (example of such a calculation in Ar-66,); but such a calculation is not necessary if the goal is simply to get the equation of motion of the electron as an entity.

Unfortunately, Poincaré doesn't write his equations in this elegant and modern way. He uses units such that $W'/c^2$ is equal to one (and therefore, he possibly misses the discovery of the mass energy relation...!), and he uses cumbersome notations which tend to obscure the content of the equations [38].

Poincaré doesn't make an explicit use of this equation. Instead of that, he controls once more that the equation is relativistically covariant and that Lorentz's approach is the only one which can give such a covariance. In the latter derivation, he extends the validity of the dynamical equation to arbitrary forces provide they transform under Lorentz's transformation as does the electromagnetic Lorentz's force. Equation (31) corresponds to the motion of one electron in some external field, neglecting the loss of energy due to its own radiation (quasi stationary motion). In § 8 , Poincaré shows that it can be extended to a system of electrons, where the motion of each electron is submitted to a common external field <u>and</u> to the electromagnetic field created by all the other ones.

At the end of his long paper, Poincaré tries to build a model of a relativistic gravitation (§9). We retain of this paragraph that it was the occasion to discover several "four partners" which transform alike "time and space co-ordinates", and also several kinematical invariants. He finally obtains a gravitational attraction law between two massive bodies which "improves" Newton's law in the sense that it is "retarded" and "relativistic covariant":

"Nous voyons d'abord que l'attraction corrigée se compose de deux composantes; l'une parallèle au vecteur qui joint les positions des deux corps, l'autre parallèle à la vitesse du corps attirant. Rappelons que quand nous parlons de la position ou de la vitesse du corps attirant, il s'agit de sa position ou de sa vitesse au moment où l'onde gravifique le quitte; pour le corps attiré, au contraire, il s'agit de sa position ou de sa vitesse du moment où l'onde gravifique l'atteint, cette onde étant supposée se propager à la vitesse de la lumière."

Poincaré notices that his tentative approach to build a new gravitational law compatible with the general requirements of Lorentz's invariance and the Newton's law as low velocity limit, cannot have a unique solution. As an example, he immediately proposes several alternative ones. These new gravitational laws were never really applied in astronomy. The sole application was an early estimate of the advance of Mercury's perihelion; it was found to be in the good direction, but too small (7'' instead of 38'', mentioned in [Po-53]). These attempts to build a gravitational law in the framework of special relativity are now totally superseded by Einstein's theory of gravitation (1913-1916).

---

[37] W' is as before the total energy of the ideal electron (electron at rest); it contains now the electrostatic energy and the work of the pressure.

[38] It might be possible that to-day hastened readers would not recognise that Poincaré's eq. 5 of § 7 is nothing else than eq. 31 here above!

As I said at the beginning of this Part 3, one should include in Einstein's 1905 work, the paper "*Ist die Trägheit eines Körpers von seinem Energie inhalt abhängig?*" that he wrote in September [Ei-05; 5]. He imagines that a body at rest in some inertial frame sends plane waves of light in some direction with an energy L/2, and simultaneously just the same quantity of light in the opposite direction [39]. Because of the symmetry of the process, the state of motion of the emitter is not changed: the body remains at rest, but its total energy is reduced by an amount L. Einstein considers then the same process from another inertial frame in uniform translation along the common x axis. In this frame, the energies of the plane waves of light have the transformed values that were computed in his preceding paper:

$$(32) \qquad \frac{L'}{2} = \frac{L}{2} \; \frac{1 \pm v \cos \phi}{\sqrt{1 - (v/c)^2}} \; ,$$

where $\phi$ is the angle of emission with respect to the x axis. Here again, because of the symmetry of the emission, the state of motion of the body is not changed, but its total energy is reduced by an amount $L/(1-(v/c)^2)^{1/2}$ . Let $E_0$ and $E_1$ be the total energies before and after the process in the rest frame, similarly $H_0$ and $H_1$ the corresponding energies in the moving frame; one has evidently:

$$(33) \qquad ( H_0 - E_0 ) - ( H_1 - E_1 ) = L \left[ \frac{1}{\sqrt{1 - ( v / c )^2}} - 1 \right] \; ;$$

Einstein envisages the possibility that in each frame a conventional additive constant could be added to the kinetic energy K in order to define the total energy : E = K + C, similarly, H = K' + D (remember that in Eq. 26 he only obtains the transformation law of the kinetic energy !). However, these conventional constants C and D do not change in the physical process, and therefore, they disappear from Eq. 33 which becomes the difference of kinetic energy (as seen from the moving frame) when light is emitted:

$$(33') \qquad K_0 - K_1 = L \left[ \frac{1}{\sqrt{1 - ( v / c )^2}} - 1 \right] \; .$$

If we compare with the expression (26) of the kinetic energy, we see that it looks as if the body had lost a fraction $L/c^2$ of its mass when emitting (at rest) an energy L in the form of radiation. Einstein concludes:

" The fact that the energy withdrawn from the body becomes energy of radiation evidently makes no difference, so that we are led to the more general conclusion that:
The mass of a body is a measure of its energy-content; if the energy changes by L, the mass changes in the same sense by L/9 x 10 [20], the energy being measured in ergs, and the mass in grammes.
It is not impossible that with bodies whose energy-content is variable to a high degree (e.g. with radium salts) the theory may be successfully put to the tests.
If the theory corresponds to the facts, radiation conveys inertia between the emitting and absorbing bodies."

---

[39] This reasoning is typical of a thermodynamics way of thinking. One considers a process, without questioning about the way it can happen: by means of which mechanism can such a localised "body" emit such plane waves? Which kind of body do we have to consider? Certainly not an electron since it cannot emit light when it remains at rest ! However, the crucial formula (26) giving the kinetic energy (which will be used at the end of the reasoning) was derived for an electron, and only boldly extended to any ponderable matter!

During the period 1906-1907, Einstein published three other proofs of his mass energy relation; this underlines enough the importance he attached to his equivalence law. May be did he also consider that this first proof was not convincing enough in view of the enormous importance of the discovery?

# Part 4.- Conclusions.

Historians sometimes like to speculate on "What would have happened if ...?" [40]. This of course is not really "good" history; but in our case, it is certainly a valid alternative to the vain quarrels on the question of priority, or on the question of a "real understanding" of the meaning of relativity by one among its creators. Let us therefore try to play the game.

The question "What would have happened if ...?" cannot reasonably be posed in the case of works which continuously developed on many years, like Lorentz's electrodynamics. In thirty years (1875-1905), the great physicist built an Oeuvre that contains so many results that profoundly influenced the development of physics that it becomes nearly impossible to envisage the possibility of its non existence.

The question "What would have happened if ...?" can more reasonably be posed in case of rather isolated events (like Poincaré's 1905 papers), or a fortiori, in case of totally unexpected events (like Einstein's original paper *"Zur Elektrodynamik bewegter Körper"* ).

"What would have happened if Poincaré's papers of 1905 did not exist"? The answer is immediate since these papers were nearly forgotten [41] and didn't really influence the later development of physics!

"What would have happened if Einstein's original paper *"Zur Elektrodynamik bewegter Körper"* of 1905 did not exist"? This time, the answer is not so easy to give. On the one hand, all important formulae existed already or would have appeared at the same time in Poincaré's papers (even if some of them are there derived in a different way; see f.i. the formulae of the new dynamics). On the other hand, it is clear that Einstein's radical approach announced some kind of new physics. Although it is most probably true that Einstein grasped some ideas in earlier works of Lorentz and Poincaré (f.i.: the idea of synchronisation of distant clocks by exchange of light signals that he probably met in [Po-00], or Lorentz's transformation formulae that he knew to exist and which played probably the role of a guideline [42]), their new derivation opened the way to further progress. This potentiality was probably perceived early by some very influent physicists (like Max Planck) who emphasised the importance of the work of the young physicist and encouraged further developments.

Einstein versus Poincaré ! We meet here one of the mysteries of scientific fate. A priori, in 1905, the comparative chances of success of the two works looked rather unequal. On the one hand, one of the greatest mathematicians of the turn of the century, whose reputation as a physicist was also great, of which each new contribution to both disciplines was eagerly looked for by the scientific world. On the other hand, a young physicist nearly unknown. At the same time, they both publish the solution of an important problem that troubled physics for many years. Of course, as we have seen, the two papers are not identical and each of them has its own merits. Among these respective merits, one has to acknowledge a higher mathematical level of reasoning in Poincaré's work, and a new insight on the problem that opens the way to further developments in Einstein's work. As an example, I recall that Poincaré keeps on firmly, with full mathematical rigour, a pure wave theory of light (as it was commonly

---

[40] For example "What would have happened if the issue of some rather decisive battle, like the battle of Waterloo, had been different ?".

[41] Except for details like Poincaré's pressure.

[42] Not necessary to insist on the fact that what I write here is pure speculation, since Einstein not only makes no reference to previous works, but denies to have known about them (see Mi-81, Pa-82).

admitted at the time), maintaining thereby the very existence of an ether whose physical properties determine those of light and gravitation [Po-06]:

"Si la propagation de l'attraction se fait avec la vitesse de la lumière, cela ne peut être par une rencontre fortuite, cela doit être parce que c'est une fonction de l'éther; et alors, il faudra chercher à pénétrer la nature de cette fonction, et la rattacher aux autres fonctions du fluide."

Einstein rejects radically the existence of such a medium and finds skilfully his way between contradictory conceptions of light, using the rather vague concept of "light complex". For Einstein, Relativity should be considered independently of Maxwell-Lorentz's electromagnetic theory; it is a fundamental Principle that defines a general framework where theories have to be developed.

This difference of conception is certainly not enough to justify that only the work of the latter is to-day considered as the founding paper of Relativity, while the work of the former is nearly forgotten. So what? Why does the living history of our century ( i.e. education, scientists knowledge and even public knowledge) remember only one name? One can put forward a lot of hypotheses: Poincaré's wrong choice when addressing his main paper to a mathematical journal not so well known to physicists (?), mathematical presentation unfamiliar to physicists (?), impact of the immediate support to Einstein's work by Planck and other important German physicists (?), early death of Poincaré (1912) which prevents a possible sharing of a Nobel Prize for Relativity with Einstein and Lorentz (?). I don't think that any one of these hypotheses can give a satisfactory explanation. It is also often said that Einstein was more "relativist" than Poincaré. The detailed comparison of the contents of the 1905 papers does not support this assertion. It is however true that an important difference of attitude with respect to the new theory exists between the two men after 1905:

- Poincaré seems to consider that the question is settled and makes no new effort to go further;

- at the opposite, 1905 is for Einstein the starting point for new, advanced studies on the consequences of the Principle of Relativity.

In this sense, Einstein is a "more relativistically committed" scientist than Poincaré. Even if one limits the discussion to special relativity (as I do in this paper), Einstein's "post 1905" contributions to relativity are important, and it is worth to stress that they were directly inspired by the new spirit he introduced in his first paper. His 1907 review paper [Ei-07] contains not only some essential clarifications on the Principles he makes use of (nl. his concepts of "identical" clocks and solids rods), some improvements of his earlier results (nl. the new relativistic mechanics), but it contains also new and important contributions: further clarification of the mass-energy relation, extension of the Principle of Relativity to accelerated frames, Principle of "Equality of the inertial and the gravitational masses" , and the first predictions concerning the influence of a gravitational field on light. If one includes General Relativity in the discussion, then of course Einstein's work is much more important than the work of any one else.

Nevertheless, in spite of this incontestable dominance, when Einstein received the Nobel Prize in Physics 1921 (given only in 1922), the mention was " for his services to Theoretical Physics, and especially for his discovery of the law of the photoelectric effect ", with *no explicit mention to his work on Relativity* [No-67]. The presentation speech was given by Professor S. Arrhenius, Chairman of the Nobel Committee for Physics. It contains only a short allusion to relativity as being the subject of lively debates in philosophical circles, which has also some astrophysical implications still under examination. Einstein could not attend the ceremony (he was then in Japan) and there was no Nobel Lecture. It was replaced by a "Lecture delivered to the Nordic Assembly of Naturalists at Gothenburg" in 1923, entitled "Fundamental ideas and problems of the theory of relativity". These facts look strange enough to awake the attention of historians and biographs. Pais makes a detailed analysis of the circumstances in §30 of his book [Pa-82] : "How Einstein got the Nobel Prize". Although many physicists nominated several times Einstein for his work on relativity, it seems that the Academy of Sweden was in no hurry to award relativity before experimental issues were clarified, first in special relativity, later in general relativity. Pais concludes that it was the Academy's bad fortune not to have anyone among its members who could competently evaluate the content of relativity theory in those early years. Leveugle proposes an alternative explanation to these strange facts [Le-94]. The Academy could have been influenced by the publication in 1921 of the obituary of Poincaré, written in 1914 by Lorentz, where the great old scientist whose authority in the matter was incontestable wrote:

> "Poincaré a formulé le Postulat de Relativité, terme qu'il a été le premier à employer".

This sentence was strong enough to stand an insurmountable obstacle to the possible attribution of a Nobel Prize for relativity to any one else. Whatever the reason, one can only notice that no Nobel Prize of Physics was attributed for one of the major discovery in physics of the century!

Let us now come back again on the question of the ether. It is generally asserted that Einstein rendered this notion completely obsolete. It is at least the conclusion that some physicists promptly derive from his 1905 introduction:

> "The introduction of a "luminiferous ether" will prove to be superfluous inasmuch as the view here to be developed will not require an "absolutely stationary space" provided with special properties, nor assign a velocity-vector to a point of the empty space in which electromagnetic processes take place".

This conclusion must however be tempered by the following other Einstein's formulation [Ei-20]:

> "According to the theory of general relativity, space possesses physical properties; therefore, in this sense, an ether exists. According to the theory of general relativity, space without ether is inconceivable, for non only the propagation of light would be impossible, but furthermore, there would be no possible existence for rods and clocks, and therefore also for space-time distances in a physical sense. However, this ether must no be conceived as having the property which characterises ponderable media, i.e. as being formed of parts that can be followed in time: the notion of motion cannot be applied to it".

In between, Einstein had built general relativity and understood that the physical properties of light and gravitation forced us to come back to a less radical position on the subject. In a sense, this was a kind of revenge for the two other creators of special relativity.

# Bibliography.

[Ab-02]    M. Abraham, *Dynamik des Elektrons* ,
Göttinger Nachrichten (1902) 20-41.

[Ar-66]    H. Arzeliès, *Rayonnement et dynamique du corpuscule chargé fortement accéléré.*
Gauthier-Villars, Paris, (1966).

[Ei-05; 1]    A. Einstein, *Über einen die Erzeugung und Verwandlung des Lichtes betreffenden heuristschen Gezichtspunkt* ,
Ann. d. Ph. **17** (1905) 132-148.

[Ei-05; 2]    A. Einstein, *Eine neue Bestimmungen des   Moleküldimension* ,
Ph. D. Thesis; Ann. d. Ph. **19** (1906) 289-306

[Ei-05; 3]    A. Einstein, *Über die von der molekularkinetischen Theorie der Warmte geforderde Bewegung von in ruhenden Flüssigkeiten suspendierten Teilchen* ,
Ann. d. Ph. **17** (1905) 549-560.

[Ei-05; 4]    A. Einstein, *Zur Elektrodynamik bewegter Körper*,
Ann. d. Ph., **17** (1905) 892-921;
English reprint in [So-22], French reprint in [Ei-93].

[Ei-05; 5]    A. Einstein, *Ist die  Trägheit eines Körpers von seinem Energie inhalt abhängig?,*
Ann. d. Ph., **18** (1905) 639-641;
English reprint in [So-22], French reprint in [Ei-93].

[Ei-05; 6]    A. Einstein, *Zur Theorie der Brownschen Bewegung* ,
Ann. d. Ph. **19** (1906) 371-381.

[Ei-06]    A. Einstein, *Das Prinzip von der Erhaltung der Schwerpunktsbewegungen und die Trägheit der Energie,*
Ann. d. Ph., **20** (1906) 627-633;
French reprint in [Ei-93].

[Ei-07]    A. Einstein, *Relativitätsprinzip und die aus demselben gezogenen Folgerungen,*
Jarhb. der Radioaktivität, **4** (1907) 411-462 & **5**, 98-99.
French reprint in [Ei-93].

[Ei-20]    A. Einstein, *L'éther et la théorie de la relativité.*
Lecture delivered at the Univ. of Leyde on 5 May 1920.
Gauthier-Villars, Paris, (1953).

[Ei-93]    A. Einstein, "Oeuvres choisies", 6 tomes,
Éd du Seuil et CNRS, (1993); see tome 2:"Relativités".

[Ga-32]    Galileo, *Dialogo, Giornata seconda* , (1632)
Ed. Naz. delle Opere di Galileo Galilei, Tome **7**, 212-214.

[Go-67]    S. Goldberg, *Henri Poincaré and Einstein's theory of relativity,*
Am. J. of Phys., **35**, 934-944.

[Ja-66]    M. Jammer, *The Conceptual Development of Quantum Mechanics* ,
McGraw-Hill Cy, New York, (1966).

[Ja-74]    M.Jammer, *The Philosophy of Quantum Mechanics* ,
Wiley-Interscience Publ., New York, (1974).

[Jc-75]    J.D. Jackson, *Classical Electrodynamics,*
(2nd. Ed.), J. Wiley & Sons, Inc., New-York, (1975).

[Ke-65]    G.H. Keswani, *Origin and concept of relativity,*
Brit. J. Phil. Sc., **15** (1965) 286-306, **16** 19-32.

[Ko-75]    I.Yu. Kobzarev, *Henri Poincaré's St.Louis lecture, and theoretical physics on the eve of the theory of relativity,*
Sov. Phys. Usp.,**17** (1975) 584-592.

[La-05]    P. Langevin, *Sur l'origine des radiations et l'inertie electromagnétique* ,
J. Physique et Radium, **4** (1905) 165-182.

[Le-94]    J. Leveugle, *Henri Poincaré et la relativité,*
"La jaune et la rouge", avril (1994).

| | |
|---|---|
| [Lg-95] | A.A. Logunov, *On the articles by H. Poincaré on the dynamics of the electron,* Publ. Depart. of the Joint Inst. for Nuclear Research, Dubna (1995), transl. by G. Pontecorvo. |
| [Lo-95] | H.A. Lorentz, *Versuch einer Theorie der elektrischen und optischen Erscheinungen in bewegten Körpern* , Leiden: E.J. Brill (1895); reprint in Collected Papers **5** , 1-137. |
| [Lo-02] | H.A. Lorentz, *The theory of electrons and the propagation of light* , Nobel Lecture, Dec. 1902; reprint in "Nobel Lectures: Physics", Elsevier (1967). |
| [Lo-04] | H.A. Lorentz, *Electromagnetic phenomena in a system moving with any velocity less than that of light* , Proc. R. Acad. Amsterdam **6** (1904) 809; reprint in Collected Papers **5** , 172-197. |
| [Lo-09] | H.A. Lorentz, *The theory of electrons and its applications to the phenomena of light and radiant heat* , Teubner, Leipzig (1909); reprint by Dover (1952). |
| [Ma-98] | J. Mawhin, *Henri Poincaré et les mathématiques sans oeillères* , Revue des Questions Scientifiques **169** (1998) 337-365. |
| [M-R-82/87] | J. Mehra and H.Rechenberg, *The historical development of quantum theory*, 5 tomes, Springer-Verlag, Berlin (1982/1987). |
| [Mi-08] | H. Minkowski, *Raum und Zeit,* Conf. Köln (1908), Phys. Zeit., **20** (1909) 104-111; English reprint in [So-22]. |
| [Mi-81] | A.I. Miller, *Albert Einstein's special theory of relativity. Emergence (1905) and early interpretation(1905-1911)*, Addison-Wesley Pub. Cy., New-York, (1981). |
| [Mi-94] | A.I. Miller, *Why did Poincaré not formulated special relativity in 1905?*, in "Henri Poincaré, Science et Philosophie", Ak. Verlag, Berlin, and Blanchard, Paris, (1994). |
| [No-67] | Nobel Lectures, *Physics 1901-1921* , Nobel Foundation & Elsevier, Amsterdam (1967). |
| [Pa-82] | A. Pais, *Subtle is the Lord... The science and life of Albert Einstein*, Oxford Univ. Press (1982). |
| [Pa-94] | M. Paty, *Poincaré et le principe de relativité*, in "Henri Poincaré, Science et Philosophie", Ak. Verlag, Berlin, and Blanchard, Paris, (1994). |
| [Pi-98] | Y. Pierseaux, *La " structure fine " de la théorie de la relativité restreinte,* Thèse de Doctorat, Université Libre de Bruxelles, (1998); also L'Harmattan, Paris, (1999). |
| [Pl-06] | M. Planck, *Das Prinzip der Relativitätund die Grundgleichungen der Mechanik* , Verh. D. Phys. Ges. ,**4** (1906) 136-141. |
| [Pl-07] | M. Planck, *Zur Dynamik bewegter Systeme* , Berliner Bericht, **13** (1907) 542-570; Ann. d. Physik, **26** (1907) 1-34. |
| [Pl-08] | M. Planck, *Bemerkungen zum Prinzip der Aktion und Reaktion in der allgemein Dynamik,* Verh. D. Phys. Ges., **9** (1908) 301-305. |
| [Po-95] | H. Poincaré, *À propos de la théorie de M. Larmor* , L'éclairage électrique (1895); reprint in "Oeuvre" Tome IX, 369-426. |
| [Po-98] | H. Poincaré, *La mesure du temps* , Rev. de Métaphysique et de Morale **6** (1898) 1-13. |
| [Po-00] | H. Poincaré, *La théorie de Lorentz et le principe de réaction* , Archives Néerlandaises des Sciences Exactes et Naturelles, 2ième série, **5** (1900) 252-278; reprint in "Oeuvre" Tome IX, 464-488. |
| [Po-01] | H. Poincaré, (Analyse de ses travaux scientifiques rédigée en 1901), "Oeuvre" Tome IX, 1-14, (1954). |
| [Po-04] | H. Poincaré, *Les principes de la physique-mathématique* , Conférence de St-Louis (1904); reprint in "Physics for a New Century", Am. Inst. of Phys. 281-299. |
| [Po-05] | H. Poincaré, *Sur la dynamique de l'électron*, C.R. Ac. Sc. Paris, **140**, 5 juin 1905,1504-1508; reprint in "Oeuvre" Tome IX, 489-493. |
| [Po-06] | H. Poincaré, *Sur la dynamique de l'électron*, Rend. Circ. Mat. Palermo, **21** (1906) 129-175; reprint in "Oeuvre" Tome IX, 494-586. |
| [Po-08] | H. Poincaré, "*La dynamique de l'électron* ", Revue générale des sciences pures et appliquées **19** (1908) 386-402; reprint in "Oeuvre" Tome IX, 551-586. |
| [Po-53] | H. Poincaré, *Les limites de la loi de Newton* , Lectures Notes: Fac. Sc. Paris 1906-1907; Published by J. Chazy in Bulletin Astronomique **17** (1953) 121-269. |
| [Re-99] | J. Reignier, *Éther et mouvement absolu au 19º siècle* , Revue des Questions Scientifiques **170** (1999) 261-282. |
| [Re-00] | J. Reignier, *À propos de la naissance de la relativité restreinte. Une suggestion concernant la "troisièmehypothèse" de H. Poincaré.* Bull. Acad.Roy. Belg., Cl. Sc., séance du 05/02/2000. (To appear). |
| [Ro-65] | L. Rosenfeld, *Theory of electrons* , Dover Publ. NY(1965). |
| [Sc-71] | H. Schwartz, *Poincaré's Rendiconti Paper on Relativity,* Am. J. Phys., **39** (1971) 1287-1294, **40** (1972) 862-871 & 1282-1287. |
| [St-95] | J. Stachel, *History of relativity* , in "20th century physics", Vol.1 (1995) 249-356, Am. Inst. of Phys. Press, (1995). |
| [So-22] | A.Sommerfeld, *Des Relativitätsprinzip*, Teubner, 4 th ed, (1922);English translation (1923), reprint by Dover (1960). |
| [To-71] | M.A. Tonnelat, *Histoire du principe de relativité* , Flammarion, Paris, (1971). |
| [Wi-51] | Ed. Whittaker, *A History of the Theories of Aether and Electricity*, Ed. of the Am. Inst. Phys. "The History of Modern Physics",Vol.7 (1971). |